\documentclass[a4paper,11pt]{article}
\usepackage[normalem]{ulem}
\usepackage{bm}
\usepackage{tabularx}
\usepackage{fancyhdr}
\usepackage{ulem}
\usepackage{mathtools}
\usepackage{amsmath}
\usepackage{hhline}
\usepackage{subcaption}

\usepackage{amssymb,amsfonts}
\usepackage{amsthm}
\usepackage{cite}
\usepackage{pos}

\title{Applying the Triad network representation to four-dimensional ATRG method}

\author*[a]{Yuto Sugimoto}
\author[a]{Shoichi Sasaki}

\affiliation[a]{Department of Physics, Tohoku University, Sendai 980-8578, Japan}

\emailAdd{sugimoto@nucl.phys.tohoku.ac.jp}
\emailAdd{ssasaki@nucl.phys.tohoku.ac.jp}

\abstract{Anisotropic Tensor Renormalization Group (ATRG) is a powerful algorithm for four-dimensional 
tensor network calculations. However, the larger bond dimensions are known to be difficult to
achieve in practice due to the higher computational cost. Adopting the methods of the minimally decomposed TRG and its triad prescriptions, we construct a triad representation of the four-dimensional ATRG
by decomposing the unit-cell tensor. We observe that this combining approach can significantly
improve the computational cost even with maintaining the convergence accuracy of the free energy
in the four-dimensional Ising model. In addition, we also show that a further improvement can be achieved in terms
of the computational cost when our proposed approach is implemented in parallel on GPUs.
}
\FullConference{The 41st International Symposium on Lattice Field Theory (LATTICE2024)\\
 28 July - 3 August 2024\\
Liverpool, UK\\}
\begin{document}
\newcommand{\tsp}{\vphantom{i_1(n+\hat{1})}}
\newcommand{\tspm}{\vphantom{-\overline{text}}}
\maketitle

\section{Introduction}
The tensor renormalization group (TRG) \cite{levinTensorRenormalizationGroup2007a} is known to be a powerful method for performing the real-space numerical renormalization group in 2D systems.
The TRG is the candidate for the first-principles calculation method of QCD at finite density because it is free from the sign problem.

However, it is necessary to extend the TRG to higher dimensions, since QCD is  defined on four dimension.
The higher-order tensor renormalization group (HOTRG) \cite{Xie:2012mjn} 
%arrows extention of 
can extend the TRG to higher dimensions by employing a coarse-graining procedure that renormalizes each space-time direction separately.
The HOTRG has been applied to the four-dimensional Ising model \cite{akiyamaPhaseTransitionFourdimensional2019,Akiyama:2019chk} up to the bond dimension of $\chi=13$.
However, in $d$ dimensions, the computational cost of the HOTRG is $O(\chi^{4d-1})$, which makes it difficult to achieve higher bond dimensions in practice.

Several alternative algorithms have been developed to reduce the high computational cost for the HOTRG in higher dimensions.
One such algorithm is the anisotropic TRG (ATRG)~\cite{adachiAnisotropicTensorRenormalization2020}, which 
stands out in terms of a significant reduction in computational costs,
scaling as $O(\chi^{2d+1})$.
Therefore, the ATRG, which is much less computationally expensive than the HOTRG, allows the analysis of four-dimensional systems with much larger bond dimensions. Notable applications of the ATRG to four-dimensional theories include bosonic systems \cite{Akiyama:2020ntf}, fermionic systems \cite{Akiyama:2020soe}, and discrete gauge systems \cite{Akiyama:2023hvt}.

There are other approaches, such as the Triad TRG (TTRG) \cite{kadohRenormalizationGroupTriad2019}, the minimally-decomposed TRG (MDTRG) \cite{nakayamaApplicationProjectiveTruncation2024,nakayamaRandomizedHigherorderTensor2023}, and its variant called the Triad-MDTRG.
In contrast to the ATRG ---which uses a bond-swapping technique to reduce the number of squeezes in the final contraction step, reducing computational scaling---, both the TTRG and the Triad-MDTRG adopt a different strategy: they decompose all tensors into 3-leg tensors and employ a randomized singular value decomposition (RSVD) to reduce the computational bottleneck in the contraction step. 
Furthermore, the MDTRG applies the internal line oversampling technique and the unit-cell decomposition of HOTRG, leading to a relatively more accurate approximation compared to standard local tensor decomposition methods.
As results, the computational cost of the TTRG is $O(\chi^{d+3})$, while that of the Triad-MDTRG is $O(qr^3\chi^{d+3})$ where the prefactor $qr^3$ comes from the number of the RSVD iterations and the internal-line oversampling parameter.
Both the MDTRG and the Triad-MDTRG have successfully achieved free energy calculations
consistent with the previous study done with the HOTRG in the three-dimensional Ising model \cite{nakayamaApplicationProjectiveTruncation2024,nakayamaRandomizedHigherorderTensor2023}, but no application to four-dimensional systems has been reported.

One of the main challenges in four-dimensional systems is how to effectively approach larger bond dimensions. The ATRG has succeeded in reducing computational costs, but still faces the problem of increasing cost as the bond dimension increases.
On the other hand, despite its potential, the MDTRG has not yet been
applied to four-dimensional systems. 

In this sense, it would be a powerful tool for four-dimensional systems, if the methods used in the Triad-MDTRG could be adopted to achieve faster computations while maintaining accuracy comparable to the ATRG.

In this work, we propose a new method, called Triad-ATRG, which 
introduces an oversampled triad representation in the ATRG with an appropriate decomposition of the unit-cell tensor used in the Triad-MDTRG. 

We also study how the ATRG and Triad-ATRG can be implemented in parallel on GPUs.

\section{The triad representation for the anisotropic tensor renormalization group}
\begin{figure}[b!]
    \hspace{-10pt}\includegraphics[scale=0.62]{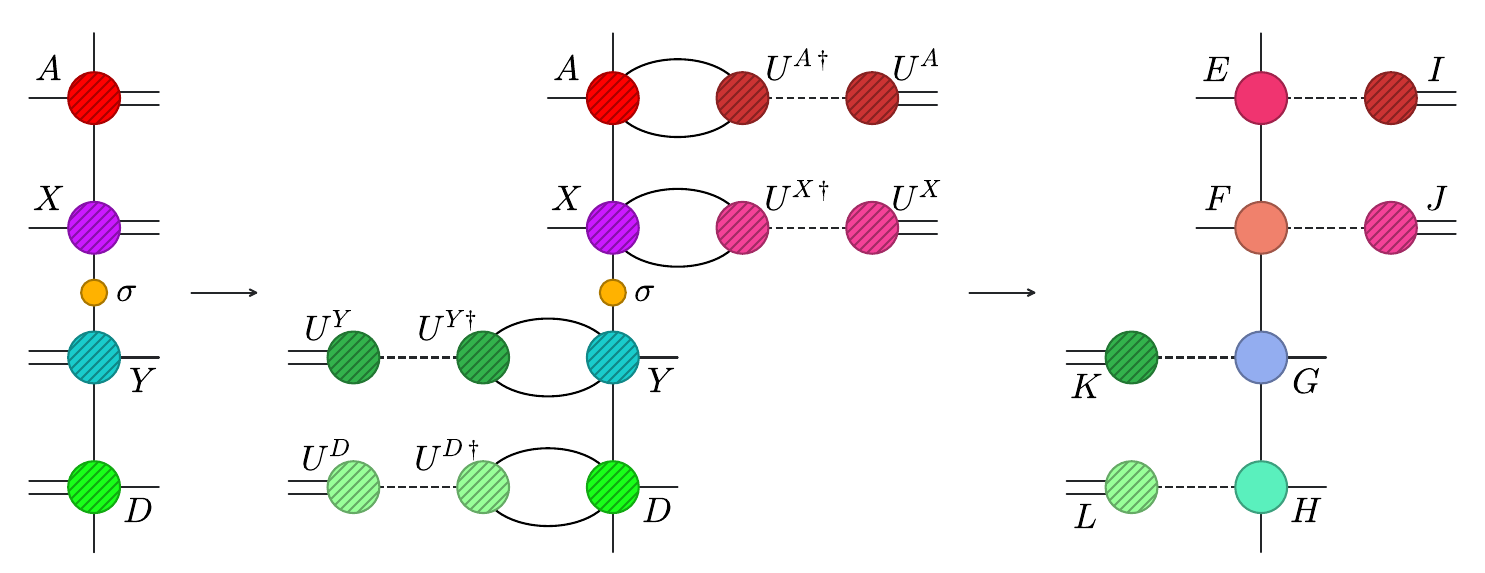}
    \caption{Schematic view of the triad representation used in the ATRG. Each dotted line is oversampled to $rD$.}\label{triadfig}
\end{figure}

In this section, let us first consider the triad representation of the ATRG, based on the methods described in Refs. \cite{nakayamaApplicationProjectiveTruncation2024,nakayamaRandomizedHigherorderTensor2023}.
We start with a unit-cell tensor of the ATRG in four dimension after bond-swapping procedure, $\Gamma=AX\sigma Y\! D $, where $A$, $X$, $Y$, $D$ are the isometries and $\sigma$ is the singular value matrix. 
Then, $\Gamma$ can be explicitly expressed as follows, 
\begin{equation}    \Gamma=\hspace{-7pt}=\sum_{\alpha,\gamma,\beta}A_{i_1(n+\hat{1})i_2(n+\hat{1})i_3(n+\hat{1})i_4(n+\hat{1})\alpha}X_{\alpha i_2(n)i_3(n) i_4(n)\vphantom{i_1(n+\hat{1})}\gamma} 
\sigma_{\gamma\gamma}Y_{\beta j_2(n+\hat{1})j_3(n+\hat{1}) j_4(n+\hat{1})\gamma}D_{j_1(n)j_2(n)j_3(n) j_4(n)\vphantom{i_1(n+\hat{1})}\beta}.
\end{equation}
where a subscript $i_\mu(n)$ denotes the bond placed on lattice site $n$ with $\hat{\mu}$ direction, while $j_\mu(n)$ denotes the opposite bond. 
A schematic figure of $\Gamma$ is shown in the left panel of Fig.~\ref{triadfig}. 
In this section, we will consider the corase-graining step along the $\hat{1}$ direction.
Let us attempt to multiply $\Gamma$ by the pairs of oversampled isometries for $A,B,C,D$ ({\it e.g.} $U^AU^{\dagger A}$ where
$U^A$ is defined for $A$ as $U^A\in\mathbb{C}^{\chi\times \chi \times r\chi}$ with the oversampling parameter $r$). These isometries are supposed to minimize the following cost function:
\begin{equation}\label{costiso}
    ||\Gamma-U^AU^{\dagger A}\Gamma||.
\end{equation}
An optimal isometry that minimizes Eq.~(\ref{costiso}) can be easily determined by the SVD of $\Gamma$.
As described in Ref.~\cite{Akiyama:2024qgv}, since $\Gamma$ is in a canonical form, to find $U^A$, we only need to consider the SVD of $AX\sigma$ as follows,
\begin{align}
    \hspace{-10pt}(AX\sigma A^\dagger X^\dagger\sigma^\dagger)_{i_2(n+\hat{1}) i_3(n+\hat{1}) \overline{i_2}(n+\hat{1}) \overline{i_3}(n+\hat{1})}\simeq\sum_{k_A}^{r\chi}U_{i_2(n+\hat{1}) i_3(n+\hat{1})k_A}^{A}\left(S_{\tsp k_Ak_A}^{A}\right)^2 U_{{\overline{i_2}(n+\hat{1})\overline{i_3}(n+\hat{1})k_A}}^{*A}.
\end{align}
The other isometries $U^B,U^X,U^D$ can be derived in the same manner. After 
multiplying by isometries as described in the center panel of Fig.~\ref{triadfig}, we get the triad representation of $\Gamma\simeq\Gamma^{\rm{triad}}$ represented by these new tensors 
as described in the right panel of Fig.~\ref{triadfig}:
\begin{align}\label{triadrepS}
    &E_{i_1(n+\hat{1})k_A i_4(n+\hat{1})\alpha}=\sum_{i_2(n+\hat{1}),i_3(n+\hat{1})}A_{i_1(n+\hat{1})i_2(n+\hat{1})i_3(n+\hat{1}) i_4(n+\hat{1})\alpha}U_{\smash[t]{i_2(n+\hat{1}) i_3(n+\hat{1})k_A}}^{*A}\\
    &F_{\alpha k_X i_4(n)\tsp\gamma}=\sum_{i_2(n),i_3(n)}X_{\alpha i_2(n)i_3(n) i_4(n)\vphantom{i_1(n+\hat{1})}\gamma}U_{\smash[t]{i_2(n) i_3(n)k_X}}^{*X}\sigma_{\gamma\gamma}\\
    &G_{\beta k_Y j_4(n+\hat{1})\gamma}=\sum_{j_2(n+\hat{1}),j_3(n+\hat{1})}Y_{\beta j_2(n+\hat{1})j_3(n+\hat{1}) j_4(n+\hat{1})\gamma}U_{\smash[t]{j_2(n+\hat{1}) j_3(n+\hat{1})k_Y}}^{*Y}\\
    &H_{j_1(n)k_D j_4(n)\tsp\beta}=\sum_{j_2(n),j_3(n)}D_{j_1(n)j_2(n)j_3(n) j_4(n)\vphantom{i_1(n+\hat{1})}\beta}U_{\smash[t]{j_2(n)j_3(n)k_D}}^{*D}
    \\
    &I_{i_2(n+\hat{1}) i_3(n+\hat{1})k_A}=U_{\smash[t]{i_2(n+\hat{1}) i_3(n+\hat{1})k_A}}^{A}\\
    &J_{\tsp i_2(n) i_3(n)k_X}=U_{\smash[t]{i_2(n) i_3(n)k_X}}^{X}\\
    &K_{j_2(n+\hat{1}) j_3(n+\hat{1})k_Y}=U_{\smash[t]{j_2(n+\hat{1}) j_3(n+\hat{1})k_Y}}^{Y}\\
    &L_{\tsp j_2(n)j_3(n)k_D}=U_{\smash[t]{j_2(n)j_3(n)k_D}}^{D}.\label{triadrepE}
\end{align}

We should remark that in four dimensions, we don't have to convert all tensors into 3-legs because it doesn't change the computational cost. Instead, we use the form that requires the least number of additional decompositions. In this sense, we refer to the network represented by 4-leg tensors $E$, $F$, $G$, $H$ and 3-leg tensors $I$, $J$, $K$, $L$ as defined in Eqs.~(\ref{triadrepS})-(\ref{triadrepE}) and the right panel of Fig.~\ref{triadfig}, as the triad representation of the ATRG.

Next, {let us introduce} the squeezers $M^{(\mu)},N^{(\mu)} (\mu=2,3,4)$ {that} minimize the cost function,
\begin{equation}
    ||\Gamma(n)^{\rm{triad}}\Gamma(n+\hat{\mu})^{\rm{triad}}-\Gamma(n)^{\rm{triad}} M^{(\mu)}N^{(\mu)}\Gamma(n+\hat{\mu})^{\rm{triad}}||,
\end{equation}
where $\Gamma(n)^{\rm{triad}}$ denotes the approximated unit-cell Tensor placed on the lattice site $n$. These squeezers are a sort 
of improved version of the original squeezers that were used for the HOTRG and ATRG, as described in Ref.~\cite{iinoBoundaryTensorRenormalization2019}.
Since $\Gamma^{\rm{triad}}$ is no longer {in canonical form}, 
all fundamental tensors in Eqs.~(\ref{triadrepS})-(\ref{triadrepE}) must be included in to derive squeezers.

\begin{figure}[tbhp]
    \begin{center}
    \includegraphics[scale=0.60]{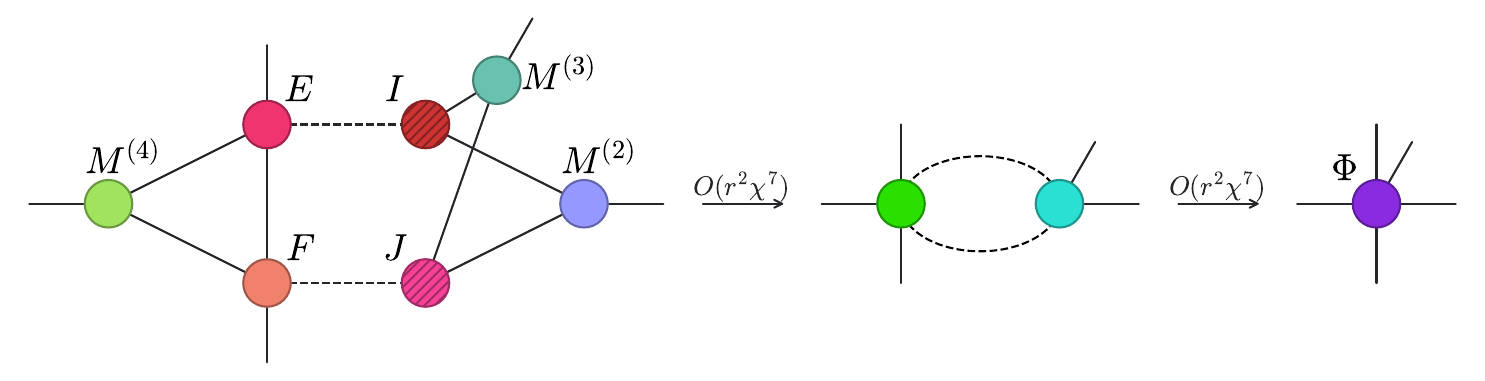}
    \caption{
    {Schematic view of the calculation process for the contraction part in} the Triad-ATRG. The bottleneck part has a cost of $O(r^2\chi^7)$}.\label{triadcont}
    \end{center}
\end{figure}

Finally, the renormalized tensors are defined by multiplying $\Gamma^{\rm{triad}}$ by the squeezers,
\begin{align}
    \label{makeMG} \MoveEqLeft[35]
    \Phi_{i_1k_2k_3 k_4\gamma}=\sum_{/\gamma,i_1,k}\left(E_{i_1(n+\hat{1})k_A i_4(n+\hat{1})\alpha}F_{\alpha k_X i_4(n)\tsp\gamma}M^{(4)}_{\smash[t]{i_4(n+\hat{1}) i_4(n)k_4}}\right)\nonumber\\
    \times\left(I_{i_2(n+\hat{1}) i_3(n+\hat{1})k_A}J_{\tsp i_2(n) i_3(n)k_X}M^{(2)}_{\smash[t]{i_2(n+\hat{1}) i_2(n)k_2}}M^{(3)}_{\smash[t]{i_3(n+\hat{1}) i_3(n)k_3}}\right) \\
    \label{makeMH} \MoveEqLeft[35] \Psi_{\tsp j_1k_2'k_3' k_4' \gamma}=\sum_{/\gamma,j_1,k'}\left(G_{\beta k_Y j_4(n+\hat{1})\gamma}H_{j_1(n)k_D j_4(n)\tsp\beta}N^{(4)}_{\smash[t]{j_4(n) j_4(n+\hat{1})k_4'}}\right)\nonumber\\
    \times\left(K_{j_2(n+\hat{1}) j_3(n+\hat{1})k_Y}L_{\tsp j_2(n)j_3(n)k_D}N^{(2)}_{\smash[t]{j_2(n) j_2(n+\hat{1})k_2'}}N^{(3)}_{\smash[t]{j_3(n) j_3(n+\hat{1})k_3'}}\right).
\end{align}
We use the notation $/\gamma,i_1,k$ in the sense of summing except for $\gamma,i_1,k$.
This procedure is the bottleneck part in the Triad-ATRG, but thanks to the Triad form, the computational cost for the contraction is reduced from $O(\chi^9)$ in the original ATRG to $O(r^2\chi^7)$.
The schematic figure of the calculation process for the contraction part is shown in Fig.~\ref{triadcont}.

\begin{figure}[t!]
    \begin{center}
            \centering
            \includegraphics[scale=0.15]{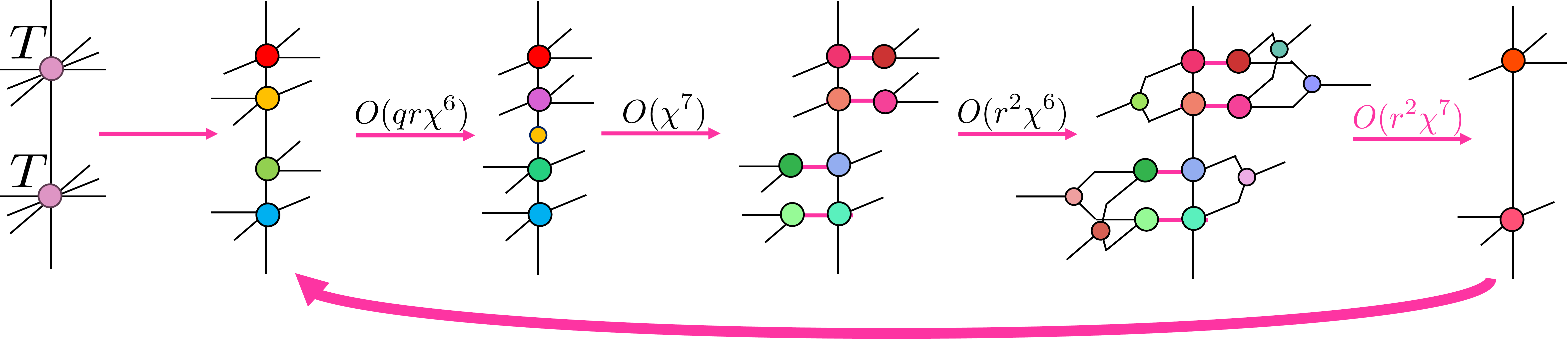}
            \caption{Schematic overview of the Triad-ATRG algorithm.}
            \label{flow}
    \end{center}
\end{figure}

We summarize the comparison of the computational costs between the ATRG and Triad-ATRG 
in Table \ref{cost} and depict a graphical representation of the Triad-ATRG algorithm in Fig.~\ref{flow}. Here, it should be noted that if $r$ is taken large, the contraction step may become a bottleneck.
It is worth pointing out that when $r = \chi$, the Triad-ATRG algorithm corresponds to the ATRG, implying that it cannot achieve higher accuracy than the ATRG.

\begin{table}[h!]

    \caption{Comparison of computational costs between the ATRG and Triad-ATRG. $q$ is the number of iterations of the QR decomposition in the RSVD}\label{cost}
    \centering
\begin{tabular}{|c|c|c|}
    \hline
    Step & ATRG & Triad ATRG\\
    \hhline{|=|=|=|}
    Bond swapping& $O(qr\chi^6)$ &$O(qr\chi^6)$\\
    \hline
    Make Triad rep.&None& $O(\chi^7)$\\
    \hline
    Squeezer step& $O(\chi^7)$ & $O(\min(\chi^7,r^2\chi^6))$\\
    \hline
    Contraction& $O(\chi^9)$ & $O(r^2\chi^7)$\\
    \hline
\end{tabular}
\end{table}

\section{Numerical results}

We apply the Triad-ATRG method to the four-dimensional Ising model.
The partition function of the Ising model can be expressed as the trace of tensor-network,
\begin{equation}
    Z=\mathrm{{tTr}} \prod_n T_{i_1(n)i_2(n)i_3(n)i_4(n)j_1(n)j_2(n)j_3(n)j_4(n)},
\end{equation}
where $T$ is the initial tensor given by
\begin{align}
    &T_{i_1(n)i_2(n)i_3(n)i_4(n)j_1(n)j_2(n)j_3(n)j_4(n)} \cr
    &=\sum_{a=1}^2 W_{ai_1(n)}W_{ai_2(n)}W_{ai_3(n)}W_{ai_4(n)}W_{aj_1(n)}W_{aj_2(n)}W_{aj_3(n)}W_{aj_4(n)}.
\end{align}
{The matrix $W$ is defined by}
\begin{equation}
    W=\begin{pmatrix}
        \cosh{\mathit{\beta}} & \sinh{\mathit{\beta}} \\
        \cosh{\mathit{\beta}} & -\sinh{\mathit{\beta}} \\
\end{pmatrix}
\end{equation}
with the inverse temperature $\beta$.
% \subsection{Free energy}

\begin{figure}[t!]
    \begin{center}
            \centering
            \includegraphics[scale=0.68]{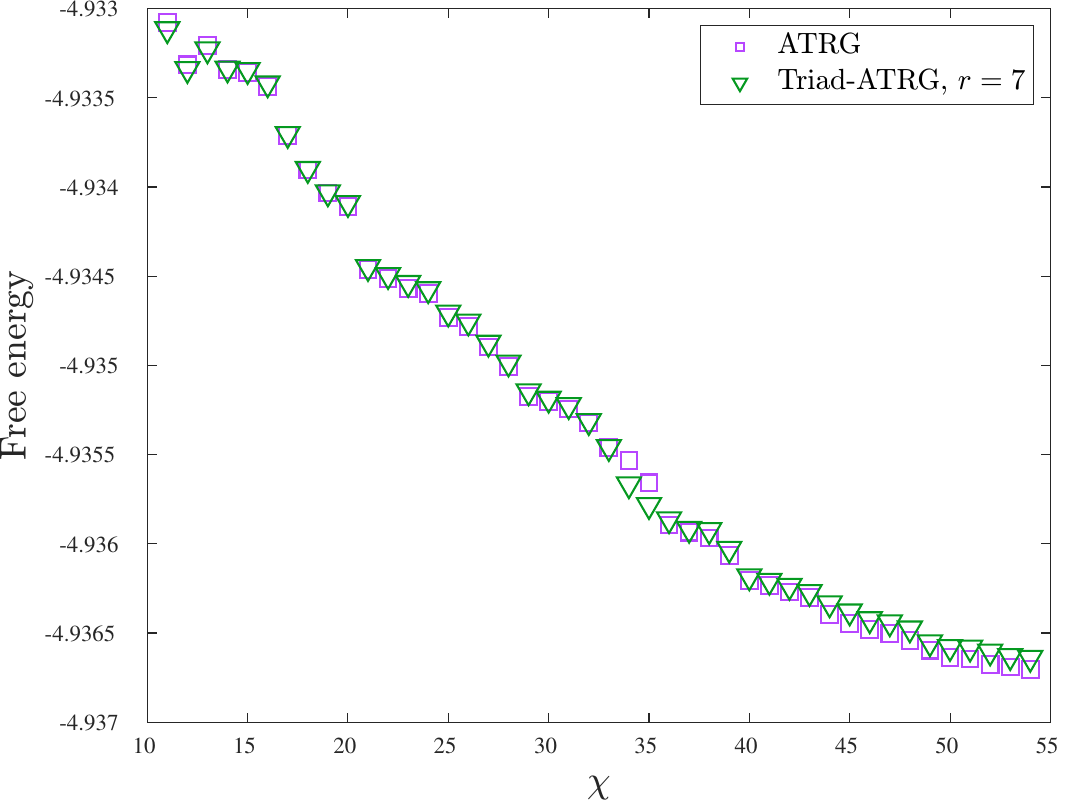}
            \caption{The approximated free energy of four-dimensional Ising model by the ATRG and Triad-ATRG with $r=7$.}
            s\label{free}
    \end{center}
\end{figure}

First of all, the free energies obtained by the Triad-ATRG and ATRG are compared.
%We first compare the free energy between the Triad-ATRG and ATRG. 
Figure~\ref{free} shows the approximated free energies obtained by using the ATRG and Triad-ATRG as a functions of bond dimension $\chi$. 
The obtained values of the approximated free energy are tabulated in Table~\ref{suutifree}. For the Triad-ATRG, the oversampling parameter is chosen to be $r=7$.
\begin{table}[h!]
    \centering
    \begin{tabular}{|c|c|c|}
    \hline
    $\chi$ & ATRG & Triad-ATRG, $r=7$  \\
    \hhline{|=|=|=|}
    38 & -4.9359676 & -4.9359235\\
    \hline
    40 & -4.9362060 & -4.9361825\\
    \hline
    42 & -4.9362695 & -4.9362360\\
    \hline
    44 & -4.9363974 & -4.9363340\\
    \hline
    46 & -4.9364809 & -4.9364227\\
    \hline
    48 & -4.9365426 & -4.9364745\\
    \hline
    50 & -4.9366373 & -4.9365787\\
    \hline
    52 & -4.9366769 & -4.9366039\\
    \hline
    54 & -4.9367035 & -4.9366392\\
    \hline
    \end{tabular}
    \caption{The free energy of the ATRG and Triad-ATRG at $r=7$.}\label{suutifree}
\end{table}
In Fig.~\ref{free},
the Triad-ATRG and ATRG methods exhibit similar convergence behavior across the whole range of bond dimension.
At $\chi=54$, the difference between the Triad-ATRG and ATRG. 
%the deviation of the Triad-ATRG from the ATRG 
was only 0.0013\%. 
Therefore, the Triad-ATRG results can reproduce the ATRG results with sufficient accuracy.
This indicates that ATRG and Triad-ATG provide comparable results in the evaluation of free energy.

\begin{figure}[htbp]
    \centering
    \begin{minipage}{0.44\textwidth}
        \centering
        \includegraphics[width=\textwidth]{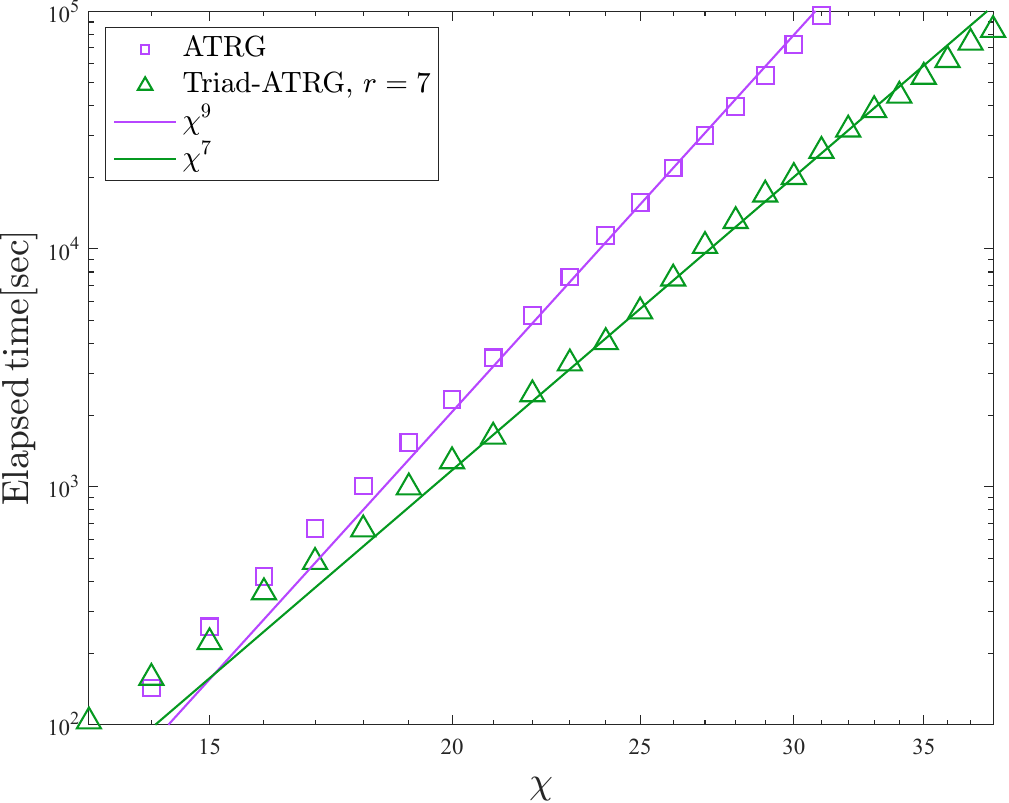}
        \caption{Scalings of computational time on a single CPU by the ATRG and Triad-ATRG with $r=7$.}
        \label{elaCPU}
    \end{minipage}
    \hfill
    \begin{minipage}{0.44\textwidth}
        \centering
        \includegraphics[width=\textwidth]{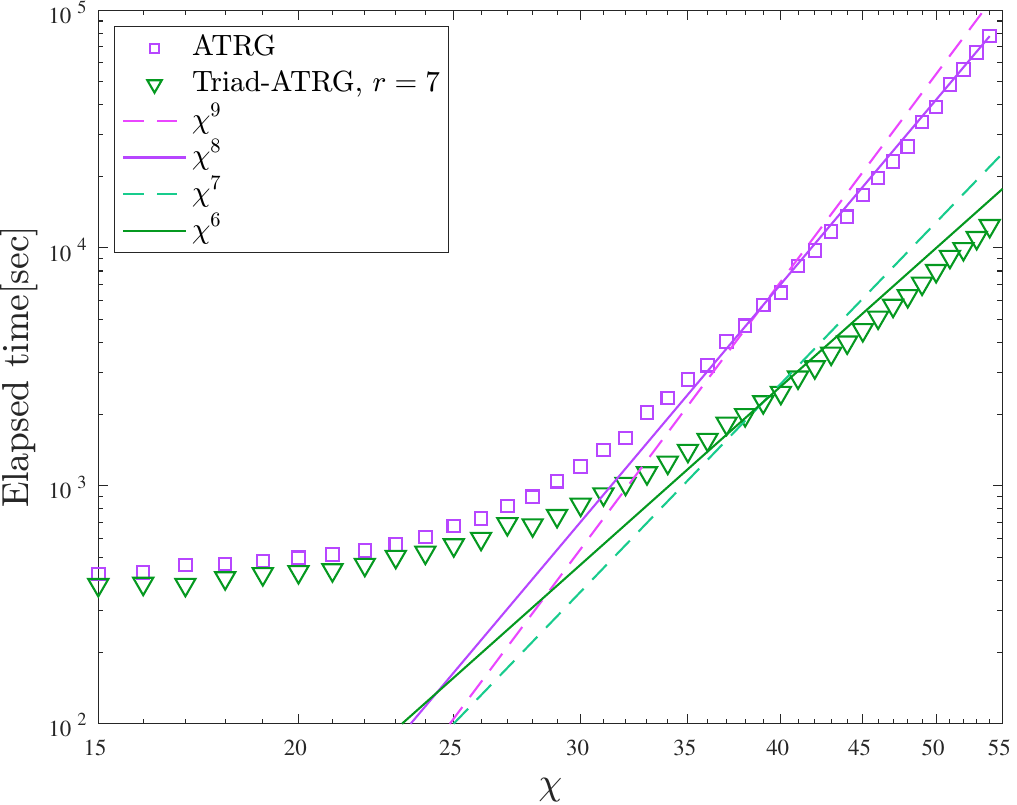}
        \caption{Scalings of computational time on two GPUs by the ATRG and Triad-ATRG with $r=7$.}
        \label{elaGPU}
    \end{minipage}
\end{figure}

We next examine the computational scaling of both methods
with respect to the bond dimension.
Figure~\ref{elaCPU} shows the computational time as a function of the bond dimension when measured on a single processor.
As shown in Table \ref{cost}, the ATRG scales as $O(\chi^7)$, while the Triad-ATRG scales as $O(\chi^9)$. 
The purple and green lines corresponding to the respective expected scaling behavior fit well with the data points 
represented by the squares (ATRG) and triangles (Triad-ATRG) in the large $\chi$ range, respectively.
This indicates that the Triad-ATRG 
has succeeded in significantly reducing costs at large bond dimensions.

GPU parallelization algorithm~\cite{jhaGPUAccelerationTensorRenormalization2024} can be applied to both methods.
In this study, two GPUs are used, and the resulting scaling of the computational time is shown in Fig.~\ref{elaGPU}.
We observe that the ATRG scales as $O(\chi^8)$ (purple solid line), while the Triad-ATRG scales as smaller than $O(\chi^6)$ (green solid line). This indicates that the Triad-ATRG could be more efficient with larger bond dimensions in GPU parallel computing.

Finally we investigate the $\chi$ and $r$ dependences of the phase transition point. We use the order parameter $X$ to determine the transition point $T_c$, which
can be effectively defined by counting the degeneracy of the ground state as discussed in Ref.~\cite{guTensorEntanglementFilteringRenormalizationApproach2009}. The definition of $X$ is given by
\[
    X^{(m)}=\frac{(\mathrm{Tr}A^{(m)})^2}{\mathrm{Tr}(A^{(m)})^2}\;\;\;\text{with }\;\;A^{(m)}_{kl}=\sum_{i_1,i_2,i_3} T_{i_1 i_2 i_3 k i_1 i_2 i_3 l}^{(m)},
\]
where $T^{(m)}$ denotes the $m$-th coarse-grained tensor.
\begin{figure}[t]
    \centering
    \includegraphics[scale=0.5]{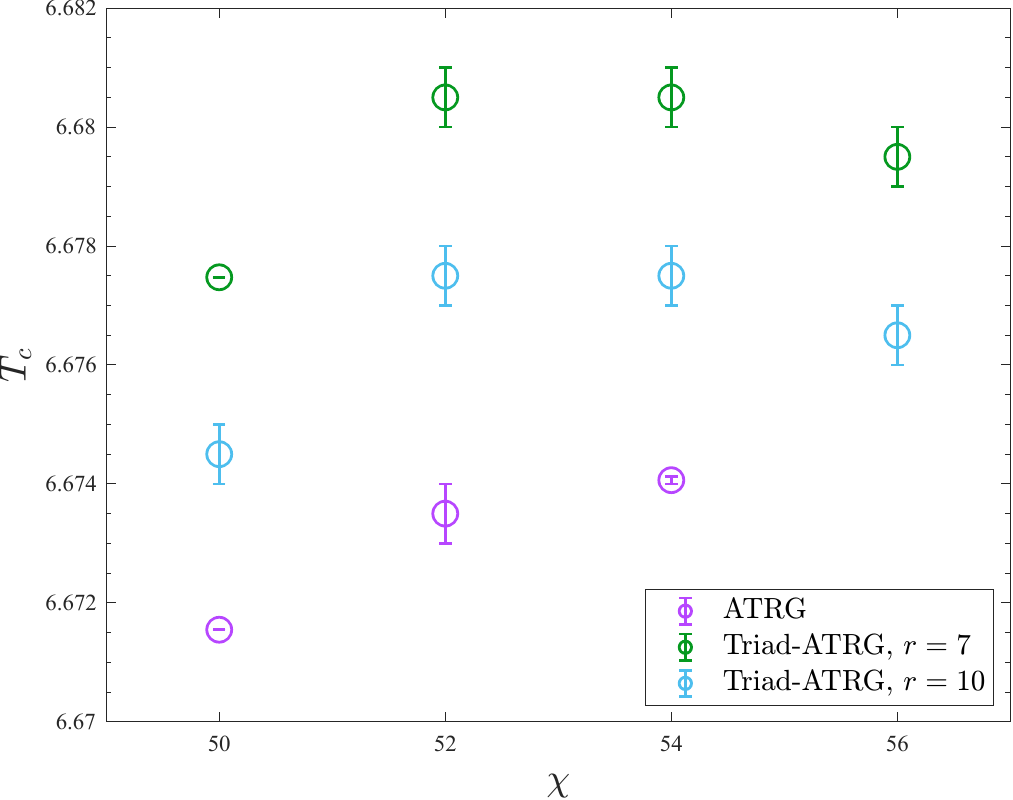}
    \caption{The phase transition point of the four-dimensional Ising model by the the ATRG and Triad-ATRG.}
    \label{phase}
\end{figure}
Figure \ref{phase} shows the phase transition point of the four-dimensional Ising model calculated by the ATRG and Triad-ATRG (with $r$=7 and 10) at different $\chi$. The error bars result from the resolution of the temperature for determining the phase transition point. 
In the range of bond dimensions explored in this study, it can be seen that not all results converge with respect to the bond dimensions.
The deviation of the Trid-ATRG result from the ATRG result is less than 0.1\% at $\chi=54$ for both $r=7$ and 10. Although further investigation for larger bond dimension is still needed, 
we can 
conclude that the Triad-ATRG shows great performance in four dimensions in terms of both computational cost and accuracy.

\section{Summary}
We proposed the Triad-ATRG method, which is an approximated version of the ATRG with the proper decomposition of the unit-cell tensor. We showed that the computational cost of the Triad-ATRG is $O(r^2\chi^7)$, which is {significantly} smaller than that of the ATRG, $O(\chi^9)$. 
Furthermore, it has been demonstrated that the Triad-ATRG can achieve significant reductions in computational costs for both CPU and GPU calculations.
We also showed that the approximation adopted by the new method to reduce costs does not affect the accuracy in the evaluation of the free energy and phase transition point, compared to the uncertainty due to the bond dimension.

\section{Acknowledgements}
Y.S. is supported by Graduate Program on Physics for the Universe (GP-PU), Tohoku University. Numerical calculations were carried out on the Yukawa-21 supercomputer at the Yukawa Institute Computer Facility.
% \clearpage

\providecommand{\href}[2]{#2}\begingroup\raggedright\endgroup

\end{document}